\documentclass[]{aastex631}

\newcommand\starname{LSST-DP1-O-614435753623041404}
\newcommand\altstarname{LSST-C25\_var1}

\begin{document}

\title{An outer-disk SX Phe variable star in Rubin Data Preview 1}

\author[0000-0002-3936-9628]{Jeffrey L. Carlin} 
\affiliation{NSF NOIRLab/NSF–DOE Vera C. Rubin Observatory HQ, 950 N. Cherry Ave., Tucson, AZ 85719, USA}

\author[0000-0001-6957-1627]{Peter S. Ferguson}
\affiliation{DiRAC Institute, Department of Astronomy, University of Washington, 3910 15th Ave NE, Seattle, WA, 98195, USA}

\author[0000-0003-4341-6172]{A. Katherina Vivas}
\affiliation{Cerro Tololo Inter-American Observatory/NSF's NOIRLab, Casilla 603, La Serena, Chile}

\author[0000-0003-3287-5250]{Neven Caplar}
\affiliation{DiRAC Institute, Department of Astronomy, University of Washington, 3910 15th Ave NE, Seattle, WA, 98195, USA}

\author[0000-0001-7179-7406]{Konstantin Malanchev}
\affil{The McWilliams Center for Cosmology \& Astrophysics, Department of Physics, Carnegie Mellon University, Pittsburgh, PA 15213, USA}

\correspondingauthor{Jeffrey L. Carlin}
\email{jcarlin@lsst.org}

\begin{abstract}

We report the discovery of an SX Phoenicis-type pulsating variable star via 217 epochs of time-series photometry from the Vera C. Rubin Observatory's Data Preview 1. The star, designated \starname~(or \altstarname~for short), has mean magnitudes of $(\langle g\rangle, \langle r\rangle) = (18.65, 18.63)$, with pulsation amplitudes of (0.60, 0.38)~mag in these bands. Its period is 0.0767 days (1.841 hours), typical of SX Phe pulsators. We derive a distance to the star of 16.6~kpc based on an SX Phe period-luminosity relation. Its position $\sim5$~kpc from the Galactic plane, in the outer Milky Way disk at a Galactocentric distance of $\sim22$~kpc, and its proper motion suggest that \altstarname~is part of the Monoceros Ring structure. This star is presented as a small taste of the many thousands of variable stars expected in Rubin/LSST data.

\end{abstract}

\section{Introduction}

The Vera C. Rubin Observatory's Legacy Survey of Space and Time (LSST; \citealt{Ivezic2019}) will unlock a vast treasure trove of deep, time-domain astronomical data. The telescope's large (8.4 meter) aperture and $\sim10~{\rm deg^2}$ field of view enable Rubin to image the entire visible sky (from Cerro Pachon in Chile) every $\sim3$ nights over a 10-year survey, building a vast time-domain dataset covering $\sim20000~{\rm deg}^2$ in the $ugrizy$ bands.

Rubin Data Preview 1 (DP1; \citealt{RTN-095}) consists of a small set of science-quality data products from images taken during commissioning of the facility. The commissioning camera, LSSTComCam, is made up of a single raft of 9 CCDs, covering a $\sim40\arcmin\times40\arcmin$ field. On-sky commissioning with LSSTComCam spanned Oct--Dec 2024; the resulting $\sim1800$ science-grade exposures were processed using the LSST Science Pipelines \citep{Bosch2018,PSTN-019} to produce DP1. The DP1 dataset \citep{10.71929/rubin/2570308} covers $\sim15~{\rm deg}^2$ of sky over 7 discrete fields. The total number of images in all $ugrizy$ bands ranges from 42 in the Fornax dSph field to 855 in the Extended Chandra Deep Field South (ECDFS).

\section{Variable stars in Rubin DP1}

DP1 data were obtained to commission the telescope and system, and have different observing cadences than expected from the LSST survey. With many visits over $<2$ months, the dataset is well-suited to searches for variable objects with short periods (e.g., $<1$ day). In this work we present the discovery of a pulsating variable star in Rubin DP1 data. We did find other variable stars in DP1,
but we did not perform a systematic search, and thus don't present a ``definitive'' catalog of DP1 variables in this paper.

We focus on the ECDFS field, which has the most observations, and the most densely-sampled time series. Our search for variables in DP1 used statistical quantities calculated over all visits (in particular over all \textit{difference} images) in which a given object should appear.\footnote{See \url{https://sdm-schemas.lsst.io/dp1.html} for the DP1 table schemas.} Statistics for each object in the \texttt{DiaObject} table (the catalog containing all objects detected in difference images) include the \texttt{StetsonJ} index (a measure of correlated multi-band variability; \citealt{Stetson1996}), the \texttt{Chi$^2$} and the inter-quartile range (IQR) of \texttt{diaSource} fluxes about the mean,
plus mean fluxes from difference and direct images, with their errors. We extracted stars with mean magnitudes and colors between $18 < g < 23$ and $-0.2 < (g-r) < 0.6$ (typical of pulsating variables in the instability strip). We then applied the \texttt{scikit-learn} \href{https://scikit-learn.org/stable/modules/generated/sklearn.ensemble.IsolationForest.html}{IsolationForest} algorithm to the \texttt{StetsonJ}, \texttt{IQR}, and \texttt{Chi$^2$} values for the $g$ and $r$-bands, and selected the top 10 objects flagged as outliers. 

\begin{figure*}[t!]
\plotone{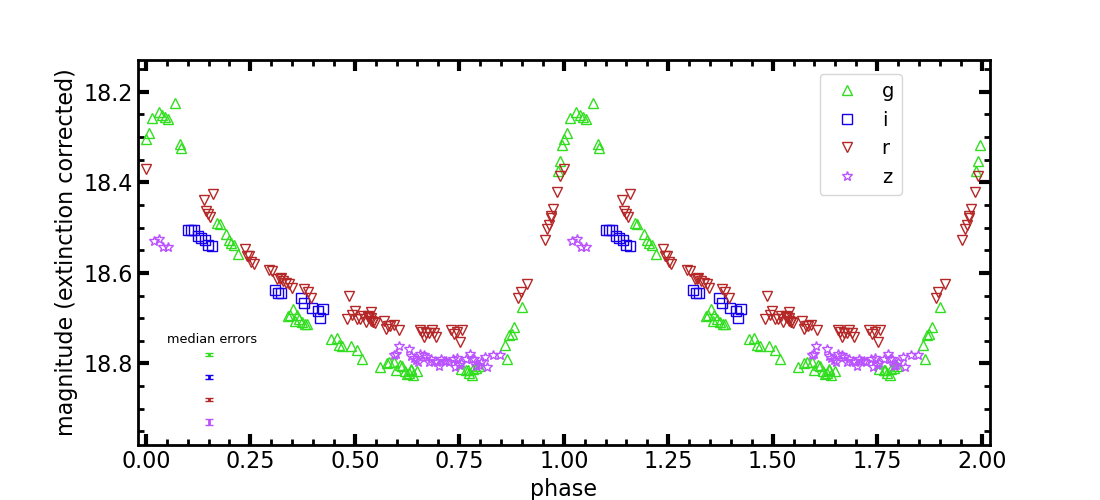}
\caption{Light curve of \altstarname~based on 197 epochs (the 20 epochs of $u$-band photometry were not included in the figure) of $griz$-band forced photometry on the visit images. The light curve has been phased with period of $0.07670835\pm0.00000267$ days.
\label{fig:lightcurve}}
\end{figure*}

Flux measurements in $ugriz$ bands were extracted from the \texttt{ForcedSourceOnDiaObject} table for the candidates, and corrected for line-of-sight extinction. 
The time-series photometry was passed to the \texttt{Psearch} period-finding software from \citet{SahaVivas2017},\footnote{Available at \url{https://github.com/AbhijitSaha/Psearch}.}
which combines the Lomb-Scargle periodogram \citep{Lomb1976, Scargle1982} and a Phase Dispersion Minimization \citep{Stellingwerf1978} technique pioneered by \citet{Lafler&Kinman1965}. Among the 10 candidates, we identify a known QSO, two known eclipsing binaries, a star flagged by Gaia as a possible multiple star, and an object classified as a galaxy by Gaia (likely showing AGN variability). Four candidates lack variability, and were flagged based on large outliers in their time series.

The remaining candidate's 217 flux measurements (20, 66, 71, 17, and 43 in $ugriz$) are well-fit by \texttt{Psearch} to a period of $0.07670835\pm0.00000267$ days (1.841 hours) and amplitudes of $A_g=0.60$ and $A_r=0.38$~mag. The phased light curve of this star, designated \starname~(hereafter \altstarname), is seen in Figure~\ref{fig:lightcurve}, showing the characteristic sawtooth shape of a pulsating variable star. The period and amplitude of \altstarname~are typical of $\delta$-Scuti ($\delta$Sct) and/or SX Phoenicis (SX Phe) type variables, which are pulsating stars below the horizontal branch. These stars become variable through different evolutionary channels. They may be main sequence stars of young/intermediate-age stellar populations ($\delta$Sct) or variable blue stragglers from old populations (SX Phe). Like other pulsating variable stars, $\delta$Sct/SX Phe are standard candles \citep[e.g.,][]{deridder23}.

A literature search finds no record of \altstarname~being identified as a pulsating variable. Its Gaia epoch photometry are unavailable in DR3, so this is likely the first identification of this star as an SX Phe-type variable.

We use the period-luminosity (P-L) relation for $\delta$Sct/SX Phe stars from \citet{Vivas2019} to estimate a distance to \altstarname~of 16.6~kpc.
A counterpart to \altstarname~is present in the Gaia DR3 \citep{GaiaDR3_2023} catalog (identifier: Gaia~DR3~2912281258855051520); its proper motion in Galactic coordinates is 
$(\mu_{l~{\rm cos}~b}, \mu_b) = (1.823, 0.536)~{\rm mas~yr}^{-1}$. Its distance combined with its position at $(l, b)=(232.72^\circ, -17.79^\circ)$ place \altstarname~at Galactocentric coordinates $(X, Y, Z) = (-17.7, -12.5, -5.0)~{\rm kpc}$ -- in the outer Galactic disk, $\sim5$~kpc below the plane. This position and its proper motion oriented in roughly the direction of disk rotation, with upward motion toward the plane, is consistent with \altstarname~being part of the Monoceros Ring structure (for a review of Monoceros, which is debated to be either a Galactic substructure or a warp of the outer disk, see \citealt{Yanny_Newberg2016}).

Because they are reliable distance indicators, and quite numerous, the vast numbers of $\delta$Sct/SX~Phe pulsating variables waiting to be found in LSST data will be valuable tracers of substructures in the disk and halo of the Milky Way.

\begin{acknowledgments}
This publication is based in part on proprietary Rubin Observatory Legacy Survey of Space and Time (LSST) data and was prepared in accordance with the Rubin Observatory data rights and access policies.
This paper makes use of LSST Science Pipelines software developed by the  \href{https://rubinobservatory.org/}{Vera C. Rubin Observatory}. We thank the Rubin Observatory for making their code available as free software at \url{https://pipelines.lsst.io}.
Support was provided by Schmidt Sciences, LLC. for N.~Caplar and K.~Malanchev.
\end{acknowledgments}

\facilities{Rubin:Simonyi (LSSTComCam), Gaia}

\bibliography{refs}{}
\bibliographystyle{aasjournalv7}

\end{document}